\documentclass{article}
\usepackage{dafx_00,amssymb,amsmath,epsfig}
\newcommand{\mm}{\vec{M}}

\title{Flexible Software Framework for Modal Synthesis}
\name{Ilia Bisnovatyi}
\address{University of Chicago \\1100 E. 58th st, \\
Chicago, IL 60637 \\
{\tt ilia@cs.uchicago.edu}}
\begin{document}
\maketitle

\begin{abstract}
Modal synthesis is an important area of physical modeling whose
exploration in the past has been held back by a large number of
control parameters, the scarcity of general-purpose design tools and
the difficulty  of obtaining the computational power required for
real-time synthesis.  This paper presents an overview of a flexible
software framework facilitating the design  and control of instruments
based on modal synthesis.  The framework is designed as a hierarchy of
polymorphic synthesis objects, representing modal structures of
various complexity.  As a method of generalizing all interactions among
the elements of a modal system, an abstract notion of {\it energy} is
introduced, and a set of energy transfer functions is provided.  Such 
abstraction leads to a design where the dynamics of interactions can
be largely separated from the specifics of particular modal structures, 
yielding an easily configurable and expandable system.  A real-time 
version of the framework has been implemented as a set of C++ classes 
along with an integrating shell and a GUI, and is currently being used to 
design and play modal instruments, as well as to survey fundamental 
properties of various modal algorithms.
\end{abstract}
\section{Introduction}
At present, the most successful work done in the area of physical
modeling has generally involved waveguide networks.  This is partly
due to the fact that for a number of synthesis algorithms, waveguide
techniques provide a large increase in computational efficiency.
However, as computer hardware gets faster and cheaper, it becomes
possible to consider other modeling representations without
compromising their generality.

Modal synthesis has been named a ``missing link'' between physical
modeling and other, more traditional synthesis techniques, such as
additive synthesis.[1]  Modal synthesis methods can range from precise
representations of a particular vibrating structure to
general sound design algorithms with large sets of control
parameters, similar in flexibility and scope to additive synthesis and
divorced from any specifics of physical models.  Synthesis based on
modal physical models can produce results theoretically 
equivalent to those of finite elements synthesis, while
allowing to approach modeling from a frequency- and energy-oriented
rather than geometry-oriented point of view, thus leading to
different kinds of data reduction and different, possibly more
intuitive, control parameters.  Unfortunately, despite the existence
of a few experimental modal systems, such as Modalys\cite{Modalys}, 
the popularity of modal synthesis is still relatively low compared to other 
synthesis methods, which can be attributed to high computational costs of a
fully realized modal system and a certain lack of numerical data for
control parameters.

Following is an overview of a software framework that has been developed 
to serve as a general-purpose design and control tool for modal synthesis
algorithms of arbitrary complexity.  This framework implements a number 
of real time modal synthesis elements and allows the user to control 
and configure the synthesis algorithms in a wide variety of ways.  Most
elements are based on a unifying abstraction of state and control 
variables, resulting in an easily expandable and configurable system.
A graphical user interface is provided to facilitate instrument design
and control in real time.

While immediately functional as a performance and compositional tool, the
framework is also currently used to systematically survey modal 
systems of varying complexity in order to determine 
optimal representations for a number of target 
phenomena\footnote{One would like to be able to generate important dynamic 
properties of sounds using the simplest possible algorithms and topologies.  
Determining exactly which properties of sounds are important is a problem 
of psycho-acoustical timbre modeling and is beyond the scope of this paper.   
For advocacy of fundamental research on sound modeling and its incorporation
into synthesis techniques, see \cite{Manifesto}.}. Among those currently 
under investigation are time- and phase-dependent response to excitation, 
saturation effects, one-way and circular energy transfer among modes, 
regions of stable and chaotic response to control parameters, and custom 
transient behavior.

\section{Synthesis Model}
\subsection{General Representation}
Modal synthesis is a technique that represents a (virtual) musical
instrument as a collection of resonant vibrating structures, each
possessing a number of {\it modes of vibration} generally associated
with particular nominal frequencies.  The time development of the
model is governed by various (possibly nonlinear) interactions among
different modes in a structure and among substructures in an instrument,
which can be used to represent the geometry, the physical properties of 
the material, and the interactions of different parts of a modeled
instrument.  Each individual mode is usually represented by a damped 
harmonic oscillator with a particular nominal frequency of vibration, while 
interactions may be based on such physical properties as resistance to 
bending, elasticity, etc.

The framework described here attempts to abstract the implementation
of all interactions from the specifics of the modal structures.  To
insure compatibility throughout the model a universal energy-like
variable \(E\) is introduced.  Each interaction between a pair of
modal  structures and/or modes is represented by an energy transfer
function (\(ETF\)), which determines the energy flow between its
arguments.   The particularities of determining the effects of energy
influx and mapping the modal state to the corresponding value of
\(E\) are left to the implementation of the modal structures themselves,
thus creating a layer of abstraction on which new objects can be easily 
built.

\subsection{Modal Structures and Nodes}
Every modal structure in the framework is implemented as a polymorphic
object, possessing a number of properties, such as its set of control
parameters \(P\), its current state \(S(t)\), and its particular instantiation 
of energy-related functions \(E_{state}\) and \(E_{feed}\).  These properties are
exposed to the user for real-time control/observation and/or to
functions that determine the interaction among modes and modal
structures.

The simplest modal structure is a modal node, which represents exactly
one normal mode of vibration.  The parameterization of a modal node is
\(P = \{m,f_0,d\}\) where \(m\) is a scaling parameter analogous to 
physical mass, \(f_0\) is a nominal frequency, and \(d\) is a decay
term.   Each of these parameters can be varied by the user at control
rate.  The state \(S\) of a modal node is comprised of a vector
\(\mm\in\mathbb{R}^{N+1}\) representing mode dynamics (i.e. the
displacement and its derivatives up to the highest order \(N\)), and
the effective frequency \(f\) and phase \(\phi\).   \(\mm\) is updated
at the audio rate (or faster if oversampling is employed), while \(f\)
and \(\phi\) are obtained from \(\mm\) via a parametrized function 
that is readjusted every cycle, reflecting the recent history of the 
mode's vibration.  For the most typical case of the standard harmonic 
oscillator, the order \(N\) is \(2\), and the highest order derivative 
is updated according to the time-discretized version of the 
standard equation of vibration:
\begin{equation}
M_2(t+1) = -dM_1(t)-\omega_0^2M_0(t),
\label{eq1}
\end{equation}
where \(\omega_0 = \frac{2\pi f_0}{srate}\), while \(M_0\) and \(M_1\) 
are updated by time-step integration of \(M_1\)
and \(M_2\) respectively, taking into account the impact of \(E_{feed}\).

\subsection{Networks of Modal Nodes}
More complex structures can be obtained by combining 
mo-dal nodes into 
{\it networks}.  A network contains a set of modal nodes and a number of
\(ETF\)s that determine the interactions among them.  A particular 
configuration of the network can be implemented at compile time for 
efficiency, or deferred to run time, if user design and micro-management
are desired.  The network exposes its set \(P\) of\  ``macro'' parameters 
to the user (i.e. \(f_0, m\), etc.) in addition to the parameters of the 
individual modal nodes contained in it.  Thus, the implementation of the
network has to translate the changes in these macro parameters into 
the corresponding changes in the parameters of the modal nodes.   
Additionally, in order to interact with other networks and modal nodes, 
a network must provide a way of calculating its state.  Finally, in cases
of certain physical models, networks need to provide a parameterization
of their states by a variable or a set of variables corresponding to the
notion of spatial location\footnote{As an example, two networks simulating
a bridge and a string may be coupled together at a particular location.  Note
that since we are treating space as an abstract parameterization, we are not 
necessarily ``limited'' to three dimensions; it becomes tempting to construct 
and listen to vibrating structures in hyper-space.}.  Typically, 
spatial parameterization of a modal network requires relatively costly 
Fourier transform, so in some cases it may be optimal to only provide such
parameterization for a particular set of coupling locations.  

Several standard network ``templates'' are included in the framework for quick 
implementation of models based on typical behaviors of various classes 
physical objects, such as strings, bars, plates, membranes, and cymbals.  
New models as well as networks with ``unrealistic'' customary behavior can be 
easily incorporated.  

\subsection{ETFs}
\(ETF\)s are responsible for all dynamic interactions in the system.  Due to 
polymorphism of the modal structures, one can use the same \(ETF\)s to describe
interactions among mo-dal nodes in a network and interactions among networks.
\(ETF\)s can generally be viewed as parametrized mappings from sets of states 
\({S_i(t)}\) to sets of energy values \(E_i\).  Different classes of \(ETF\)s
are distinguished, along the lines of the hierarchy of complexity levels
described below.  Further subdivision is generally required for efficiency -
for example, nonlinear functions can be classified by types of nonlinearity.
If the synthesis is closely based on physical models, \(ETF\)s acting on complex 
structures will be dependent on spatial parameters provided by the networks.

In addition to the \(ETF\)s whose domain and range belong to the same types of 
objects, a special case of an \(ETF\) that acts on individual modal nodes of a
network but derives its values from the macro state of the network is provided
for efficient implementation of global constraints on complex modal structures.

\subsection{Levels of Complexity}
While existing modal synthesis systems are usually closely associated 
with physical models of particular vibrating objects, the framework 
described here was intended as a more general tool, allowing 
the user to generate the largest possible range of algorithms that 
can be implemented using the modal paradigm.  In order to organize this 
space of algorithms, a classification was chosen
on the basis of complexity\footnote{Here ``complexity'' is intended to mean 
the level of sophistication of the interaction among modal objects and \(ETF\)s, 
although higher complexity in this sense generally does correspond to higher 
computational complexity.}.  Following levels are distinguished (in the order
of increased complexity):
\begin{itemize}
\item linear modal nodes and \(ETF\)s, \(f\)=\(f_0\) for all modal nodes
at all times, \(ETF\)s ignore phase information
\item phase-dependent terms are introduced into \(ETF\)s and into excitation 
response.
\item \(f\) does not necessarily equal \(f_0\), but linearity of the \(ETF\)s and
the equations of vibration is maintained.  In practical terms, this means
that \(E_{feed}\) can affect the frequency of vibration.
\item nonlinearities are allowed in \(ETF\)s and in the vibration equations for the modes.
\end{itemize}
Clearly, this is only an approximate classification which can be easily refined 
further; nevertheless, the author believes it to be a good starting point for 
investigating the kinds of qualitative phenomena that can or cannot be 
recreated by restricting the complexity of the algorithms.   Such investigation 
is the current primary focus of our work, and some preliminary results will be 
reported along with the sound examples\footnote{Most of the examples are available 
on-line at http://people.cs.uchicago.edu/\~~ilia/modal/sound\_examples/.}.

\section{Implementation}
\subsection{Synthesis}
The framework has been implemented as a set of C++ clas-ses.  An 
integrating shell and a user interface have been provided.  Special
attention has been given to maximum portability and real-time performance.  
All the non-time-critical dynamic data structures requiring heavy memory 
management have been implemented using Standard Template Library\cite{STL}, 
while for time-critical data either static memory allocation or, in rare 
cases, specialized memory management routines have been employed.  The code
is undergoing continued profiling in the hope of eliminating any unnecessary
platform-specific routines.  

A typical approach to modular software synthesis is a signal-flow
[quasi-]sequential design, where units processing data at audio rate
are chained together in a sequence.  Unfortunately, this approach is
not appropriate for a system with a complicated (potentially circular) 
topology and close 2-way and n-way integration among the elements.
Thus, a specialized scheduling algorithm had to be designed for the framework;
its general function is to look through the list of all couplings\footnote{This
list is a dynamically maintained structure, since new couplings can be created
and deleted in real time}, compute the \(ETF\)s and update the energy values 
in a way that simulates parallelism, after which, all modal structures are 
updated accordingly.  

The standard distinction between control rate and audio rate is maintained, 
with the addition of the energy-coupling rate.  The set of controls is 
subdivided into ``playable'' controls, such as fundamental frequency or mass 
values and ``system state'' controls, such as, for example, the total number 
of modal nodes.  While both sets can be updated in real time, it is assumed 
that the former corresponds to performance, while the latter --- to instrument 
design; in practical terms this means that the changes to ``playable'' 
control parameters have to be continuously integrated into the current
state of the system, while altering the ``system state'' parameters does
not necessarily maintain that continuity.

During the development of the framework, various choi-ces had to be made
with respect to maintaining a proper balance between precision and
computational  efficiency of the system.  While the mechanisms for higher 
precision have in many cases been provided, it was generally assumed that 
due to the experimental and interactive nature of the framework, 
real-time performance should be given priority as long as the qualitative 
nature of the results was not being compromised.

\subsection{User Interface}
Considering an appropriate user interface is an important part of
designing any real-time interactive application.  At the initial
stages of the  development, there was an effort to provide a front end
to the framework via an interpreted programming language, such as
Python.  However, while having advantages of flexibility and
expandability, such interface did not  appear adequate for real-time
control or data presentation.  It has since been superseded by a set of
graphical UI tools tailored to various parts of the framework.  It is
still possible to read and edit most parameters as text, since the
interface retrieves and stores its state in simply formatted text
files.   Expanding the interface with a  fully functional interpreter
is still being considered, but is not of high priority at this point.

The GUI tools provided with the framework have been written on top of
FLTK\cite{FLTK}, which was chosen for its small size and high
portability.  They include several C++ classes implementing UI
elements corresponding to different parameter sets of a modal network
and a general purpose integrating shell.  The purpose of
network-specific  classes is to provide the user with quick and easy
access to the state and parameterization of the network and the modal
nodes it contains.   As an example, in order to display and control
the \(f_0\)'s of the modal nodes in a modal network \(N\), one needs to 
execute the following code:
\begin{verbatim}
   MD_Network *N;   
   MD_FreqWindow *fwnd;
   aNet = new MD_Network("simple_net",16);
   ...
   fwnd = new MD_FreqWindow(N);
\end{verbatim}
The last line creates a frequency-slider window attached to the modal
network, similar to the example below:
\begin{figure}[ht]
\centerline{\epsfig{figure=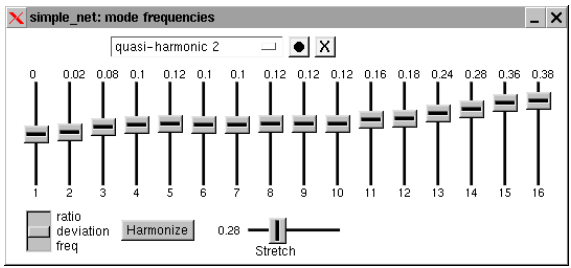,width=75mm}}
\caption{{\it An $f_0$-controlling window for a network with 16 modal nodes.}}
\label{iresp}
\end{figure}\\
In many cases, the user has several options for data representation; for example
the \(f_0\)'s of modal nodes can be represented as ratios of network's fundamental,
deviations from its harmonic partials, or arbitrary frequency values.  A system
of ``snapshots'' for control parameters has been implemented, allowing to easily
store and recall any of the user's settings, either for a window or for the
entire network/instrument.  All snapshots for a particular instrument 
are stored in human-readable files corresponding to that instrument.

The dynamic allocation of $ETF$s is facilitated by a grid window, 
representing a matrix of pairwise couplings among modal nodes:
\begin{figure}[ht]
\centerline{\epsfig{figure=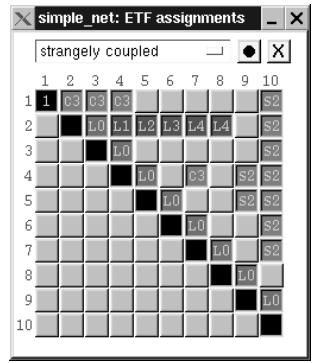,width=40mm}}
\caption{{\it An example coupling matrix for a network with 10 modal nodes.}}
\label{iresp}
\end{figure}\\
Individual cells can also be combined into groups to represent couplings 
affecting more than two nodes at a time.  \(ETF\)s for a particular cell
or a group of cells are selected from a menu containing a set of prepared 
function templates, with individual parameterization of each instance.

\section{Performance}
The framework and the UI have been developed under Linux, along with a parallel
port to Win9x.  On both platforms, real-time output to soundcard drivers (OSS
for Linux and DirectX for Windows) is provided, along with optional output to 
disk.  Partial MIDI control has been implemented, and fully customizable
integration of MIDI control surfaces with the GUI is in the works.  At the moment 
it is difficult to give a measure of performance, since a complete modal 
instrument can include a vast number of structures and \(ETF\)s with 
varying computational costs.  As a token example of performance, an AMD 
K6-2 350 Mhz machine under Linux was able to run a model of a stretched string 
with about 80 modal nodes at 44100 Hz audio rate\footnote{Due to potential 
aliasing, this estimate is meaningful only for relatively low fundamentals}.

\section{Summary}
A general framework for modal synthesis has been presented.  Its 
main strengths are its flexibility, its abstraction of dynamic 
interactions from the properties 
of modal structures, and its real-time design.  It can be used, along with the
integrating shell and GUI, as a stand-alone real-time sound design
and performance tool, or as a code base for further development of modal 
objects and algorithms.  The framework is currently utilized in a survey of a 
large set of modal algorithms of varying complexity in the hopes of providing 
new insights into theory and practice of modal synthesis.

\newpage
\bibliographystyle{IEEEbib}

\end{document}